\documentclass{aastex}

\slugcomment{accepted by Astronomical Journal (October 2000 Issue)}

\shorttitle{Carbon Star survey II.}
\shortauthors{Battinelli \& Demers}

\begin{document}


\title{Carbon Star Survey of Local Group Dwarf Galaxies. II. \\
Pegasus, DDO 210 and Tucana}


\author{Paolo Battinelli\altaffilmark{1}}
\affil{Osservatorio Astronomico di Roma, viale del Parco Mellini 84, I-00136
Roma Italy}
\email{battinel@oarhp1.rm.astro.it}
\and
\author{Serge Demers\altaffilmark{1}}
\affil{D\'epartement de Physique Universit\'e de Montr\'eal, Montreal, Qc
H3C 3J7, Canada}
\email{demers@astro.umontreal.ca}

\altaffiltext{1}{Visiting Astronomer, Cerro Tololo Inter-American Observatory.
CTIO is operated by AURA, Inc.\ under contract to the National Science
Foundation.}


\begin{abstract}
We present the latest results of our ongoing four filter photometric survey of C 
stars in Local Group dwarf irregular galaxies. Observations of the two low luminosity dwarf 
irregular galaxies, Pegasus and DDO 210, revealed  respectively 40 and 3 C 
stars,
assuming that the reddening of Pegasus is negligible. No C stars were identified
in Tucana. Our observations permit the estimation of the 
CMD contamination by foreground M dwarf thus yielding reliable C/M ratios. 
Our R, I photometry of the C stars cannot be used to solve the
extinction controversy toward Pegasus. The three C stars in DDO 210 are
quite bright when compared to C star populations in dwarf galaxies. A
larger fainter population in that galaxy seems however improbable. The 
statistics of C stars, currently on hand for dwarf galaxies, show a well-defined
trend with the absolute magnitude of galaxies.
\end{abstract}


\keywords{galaxies: individual (Pegasus) (Tucana) (DDO 210)  --- galaxies: stellar
content --- stars: carbon}

\section{INTRODUCTION}
Albert et al. (2000, Paper I) have shown the need of an homogeneous set of
C stars data for nearby galaxies if one intends to compare carbon star
 statistics of
Local Group galaxies. A flaw of previously published
surveys is the very limited area surveyed in each galaxy, thus requiring
ill defined extrapolation to assess the global numbers of C or M stars. For
example, Cook, Aaronson \& Norris (1986) surveyed only two $1.5'\times 2.5'$
areas in NGC 6822 and IC 1613, representing a few percents of the total area
of each galaxy. $320\times 512$ pixels CCDs used in the 1980's yielded few C
star candidates in external galaxies. For this reason the published statistics
are somewhat uncertain. Pritchet et al. (1987) estimated the C/M ratio of
NGC 55 from 12 C stars and 9 M stars for which two were assumed to be 
foreground dwarfs. Exception to this rule do fortunately exist. Dwarf
spheroidal galaxies such as Fornax, Leo I or Phoenix have been entirely
surveyed. 

Furthermore, different authors have adopted different color limit
for the selection of M stars, see Table 1 of Groenewegen (1999). Thus the
published C/M ratios are based on statistics for stars later than M0, M3 or
M5 making the comparison between galaxies rather difficult. 
Having the homogeneous survey as a goal,
we are then pursuing a systematic survey of Local Group dwarf galaxies
to acquire uniform data.
This second paper deals with two dwarf irregular galaxies and one dwarf
spheroidal galaxy.
Table 1 presents the currently known parameters for each galaxy. The size of
galaxies is normally based on their bright population. Faint halo of nearby
galaxies have been found to extend much farther than their Holmberg radius,
see for example the case of DDO 187 investigated by Aparicio, Tikhonov \&
Karachentsev (2000) and IC 1613 (Paper I). 

Pegasus (DDO 216) is among the faint dwarf irregular galaxies of the Local
Group. Even though it was discovered in the 1950's, it was not recognized
as a nearby galaxy and possibly a member of the Local Group until Fisher \&
Tully (1975) detected its HI and measured its radial velocity. Pegasus' 
distance has however been somewhat ill defined since its recognition as
a possible Local Group member. The first CCD investigation of Pegasus, done
by Hoessel \& Mould (1982) sets the galaxy well outside the Local Group, at a
distance of 1.7 Mpc. A few years later, Hoessel et al. (1990) confirmed this
distance using a few newly identified Cepheids. Pegasus' distance was
cut by a factor two by Aparicio (1994) who based his findings on the apparent 
I magnitude of
the tip of the red giant branch. The distance of Pegasus was further reduced
by Gallagher et al. (1998) who concluded that Pegasus is metal poor and
suffers an unusually
large amount of extinction (for an object at b = $-43^\circ$) estimated to
be A$_V$ = 0.47. As we can see however in Table 1, the adopted distance has
little effect on the calculated absolute magnitude of Pegasus. We shall see
that the number of C stars in Pegasus is a strong function of its adopted
color excess because our criterion is based on the (R--I) color of the 
candidates.  Hoessel \& Mould (1982) have identified 21 stars in Pegasus
with (B--V) $>$ 2.0. They concluded that these stars, with a $<I>$ = 21.32,
must be carbon stars.  

\placetable{tbl-1}
DDO 210, sometimes called the Aquarius system,  is a faint dwarf irregular 
galaxy which kept its original name from the
catalog of dwarf galaxies published by van den Bergh (1959). Its membership
to the Local Group was deduced from its radial velocity obtained from HI
observations by Lo, Sargent \& Young (1993) but only recently a reliable
distance was determined by Lee et al. (1999).

Tucana is one of the newly identified members of the ever increasing Local
Group family. Listed in the Southern Galaxy Catalog (Corwin et al. 1985), this
galaxy was overlooked until Lavery (1990) re-discovered it. Saviane et al.
(1996) estimated its distance to be 870 kpc. Tucana is a dwarf spheroidal
galaxy of type dE5, according to Lavery and Mighell (1992).

\section{OBSERVATIONS}
The observations presented in this paper were obtained at the CTIO 1.5 m
telescope with a 2048$\times$2048 Tek CCD during a five night run in August
1999. The telescope was employed at the f/13.5 focus, yielding a pixel size
of $0.24''$ and a field of view of $8.2'\times8.2'$. This field is deemed
suitable for the small galaxies under investigation. It further allows to
evaluate the foreground contamination. To photometrically identify C stars,
we employed the technique used in Paper I and described by Brewer et al.
(1995). Standard Kron-Cousins R and I filters are used along with CN and TiO
interference filters respectively centered at 810 nm and 770 nm. Both filters
have a width of 30 nm.  

Sky flats were obtained each night through each filters. Calibration to the
standard R, I system was done using Landolt's (1992) equatorial standards
observed during the course of the night. Several equatorial fields were
observed along with the T Phe field located at declination of $-46^\circ$.
Extinction coefficients and transformation equations were obtained by
multilinear regressions. The technique used is similar to the one detailed
by Grondin et al. (1990). Once the color term of the magnitude transformation
($\beta_i$) and the slope of the color transformation ($\beta_{ri}$) were
determined for each night, we combined the standards for the five nights to
determine one set of coefficients. This set, presented in Table 2, is used
to transform all the data. The number of standard stars used in each 
solution is given in the first column of Table 2.

\placetable{tbl-2}
We have also investigated the possible need of an atmospheric extinction
coefficient for the (CN--TiO) color index. During the course of our run,
we observed continuously, during nearly six hours, one of our program galaxy
Sgr DIG. The CN and TiO observations were analyzed in the same way than the
data presented in this paper. The slopes of the instrumental magnitudes versus
airmasses for CN and TiO are both identical and equal to $0.39\pm0.05$. 
We thus conclude that there is no need to take into account the atmospheric
extinction for the index (CN--TiO).

After the standard prereduction of trimming, bias subtraction, and sky
flat-fielding,
the photometric reductions were done by fitting model point-spread
functions (PSFs) using DAOPHOT/ALLSTAR/ALLFRAME series of programs (Stetson
1987, 1994) in the following way: we combine, using MONTAGE2, all the images
of the target irrespective of the filter to produce a deep image devoid of
cosmic rays. 
ALLSTAR was then used on this deep image to derive a list
of stellar images and produce a second image where the stars, found in the
first pass, are removed. This subtracted image is also processed through
ALLSTAR to find faint stars missed in the first pass. The second list of
stars is added to the first one. The final list is then used for the analysis
of the individual frames using ALLFRAME. This program fits model PSFs to
stellar objects in all the frames simultaneously.

The exposure times were selected to provide a S/N = 20 for stars whose 
magnitude is one magnitude fainter than the expected median magnitude of
the C stars. We set, from previous experience, the length of the CN and TiO
exposures at least three  times the I exposures. The journal of the observations
presented here, is given in Table 3. In the I wave band, C stars are among 
the brightest stars in dwarf galaxies. Thus, we do not need to explore 
several magnitudes below the red giant tip to observe them all.  For this
reason, we can survey Local Group galaxies with a medium size telescope.
\placetable{tbl-3}
\subsection{\it Comparison with published photometry}
Aparicio \& Gallart (1995) have published VRI photometry of Pegasus. The file
of photometry was kindly provided by Dr. Aparicio. We are thus in position to
compare our magnitudes and colors (BD) with theirs (AG).  We were able to match some
1200 stars common to both studies. We compute the following differences.
$$ \Delta_i = I_{BD} - I_{AG}  = -0.018 \pm 0.002$$
$$ \Delta_{ri} = (R-I)_{BD} - (R-I)_{AG}  = -0.043 \pm 0.002$$
The color difference is reduced to --0.007 if one compares only stars with
(R--I) $>$ 0.7. 
Our magnitudes are thus in excellent agreement with those of Aparicio 
\& Gallart but our overall colors may show a slight color effect relative
to their colors.
.
The Pegasus photometry of
Hoessel \& Mould (1982) is not in the same photometric system than ours,
the magnitude comparison would be quite laborious. It is also nearly
impossible to identify the stars from their published small scale photograph.

A similar test can also be done with the short table published by Lee et al.
(1999). They, however, do not quote uncertainties for their magnitudes.
Comparison of I magnitude of 26 stars, common to both investigations, yields
$\Delta_i$ = --0.02. A list of Tucana photometry was also published by 
Saviane et al. (1996). Similarly to the DDO 210 data, no uncertainties
are quoted. Furthermore most of their I magnitudes are quite faint and not
suitable to be compared with our data. We find, from the comparison of 
57 stars, $\Delta_i$ = --0.05. These three comparisons give us confidence in
our magnitudes and colors.  

\section{RESULTS}
\subsection{\it Color excess and extinction in our photometric system}
Since we used Kron-Cousins R and I filters, it is
necessary to transform published extinctions and color excesses into that
system. The relation E(R--I) = 0.82E(B--V) established by Schmidt (1976)
holds for the Johnson's R and I filters. However, from Fernie's (1983)
relationship between the Kron-Cousins (KC) (R--I) and the Johnson (R--I) for 
red stars, one sees that the E(R--I) color excess in both systems 
are nearly identical. We then take Schmidt's relation to calculate the
reddening and adopt the relationship between the extinction A$_{I_{KC}}$ and the
color excess from Dean et al. (1978): A$_{I_{KC}}$ = 2.4E(R--I).

The interstellar extinction has very little effect on the (CN -- TiO) color
index because the two wavelengths are near each other. We calculate that
E(CN -- TiO) corresponds to  4\% of the E(B--V), thus quite negligible for
the galaxies under study. For extreme cases, such as IC 10 which has nearly
one magnitude of reddening, one would have to take into account the (CN -- TiO)
reddening. 
\placefigure{fig1}
\subsection{\it Selection of C and M stars}
We adopt, for the distinction between  C and M stars, the same criteria 
employed in Paper I. The identification of C and M stars is
based on the location of the stars in the color-color diagram. 
The two zones are drawn
following the criteria of Paper I. The R--I = 0.90 blue limit follows
Brewer et al. (1995) original definition. It  applies however 
to a reddening free color-color diagram.
The (CN--TiO) index is not expected to vary continuously from M giants
to C stars. Carbon suddenly have bands not seen in M stars. The intermediate
S stars, with weak CN and TiO bands, are very rare. They will have a (CN--TiO)
even more negative than the M stars.
Furthermore, 
C stars are not expected to define a narrow region in the color-color diagram
(Cook \& Aaronson 1989).
This explains why a larger scatter is observed for C stars than for M stars.

The color-color diagram of Pegasus is shown in Figure 1. Here we plot only
stars for which $\sigma_{R-I} < 0.1$. This essentially corresponds to stars
brighter than I $\approx$ 21.25 and with $\sigma_{CN-TiO} < 0.16$. Relaxing
this condition would increase the number of fainter C star candidates but
we believe that the reality of their nature would be questionable. 
  
Our observations reveal that
there are some forty C stars in Pegasus, three C stars 
in DDO 210 and none in Tucana. The color-color diagrams
obtained for DDO 210 and Tucana are presented respectively in Figure 2 and
Figure 3. The controversy relative to the extinction toward Pegasus compels
us to take into account the possibility that  its extinction could be
unusually hight. 
We discuss this alternative in the next section.
\placefigure{fig2}
\placefigure{fig3}
\placetable{tbl-4}
The 40 C star candidates in Pegasus, with
(R--I) $>$ 0.92 are listed in Table 4. The J2000 equatorial coordinates 
were established
using ESO's Skycat facility and the IRAF tasks CCMAP and CCTRAN. 
I magnitudes and colors, along with their uncertainties, are presented.
Note that this set of C stars is selected by assuming that 
E(R--I) = 0.02. If the color excess is larger, then one would have
to move the blue limit of (R--I) to a redder value. 
Similar data are presented for the  three C stars found in DDO 210 
in Table 5. 
\placetable{tbl-5}
\section{DISCUSSION}
\subsection{\it The evaluation of foreground M stars}
The galaxies selected for our CTIO observations are intentionally small for the
CCD field. We can then use the distribution of M stars in the periphery to
evaluate the foreground contribution to the total number of M stars. The
worst case for this evaluation is Pegasus because its C star distribution
occupies about 70\% of the CCD field.  
C and M stars of the Pegasus field are plotted
on Figure 4. The galaxy extends diagonally across the field. We select the 
two corner areas, defined by the dotted lines, as representative of the
foreground M star population. The number of M stars, to be compared with
the number of C stars, is given in Table 6 where N$_{M_{fg}}$ refers to 
the number of foreground stars to be deleted from the total number of M
stars, thus yielding a true C/M ratio. For DDO 210, which is aligned E -- W,
the foreground M stars are evaluated from strips north and south of
the galaxy.
\placefigure{fig4}
\placefigure{fig5}
\placetable{tbl-6}
The magnitude distribution of M stars in the Pegasus field is compared
to the distribution of the foreground M stars in the upper panel of Figure 5. 
Essentially, all M stars brighter than $I\approx 19.6$ 
are foreground stars. The lower panel of Figure 5 reveals that there is
no M stars in Pegasus redder than R--I = 1.6, corresponding to spectral type
M5 (Th\'e et al. 1984). The same limit was found in IC 1613 (Paper I).
\subsection{\it The number of C stars and the M$_V$ of the parent galaxy}
The number of C stars in a galaxy appears to be function of its mass and
also of its metal abundance. Such relationships have been presented recently
by Groenewegen (1999). We present, in the top panel of Figure 6, what we
believe to be the most reliable log(N$_c$) vs M$_V$ relationship. Only galaxies
which have been fully surveyed are presented, this condition excludes the
Magellanic Clouds. Dwarf spheroidals with blue C stars
are excluded: only Fornax (104 C stars) and Leo I (16 C stars) remain. 
According to this relation, galaxies brighter than M$_V$ = $-15$ are 
expected to contain several hundred C stars. Excluding the
Magellanic Clouds, at least half dozen Local Group dwarfs are as bright as
this magnitude. It will eventually be interesting to compare the number of C stars
in galaxies of the same absolute magnitude but of different metallicity.
NGC 147 and NGC 185, two companions of M31, offer this possibility. 

\subsection{\it C stars and the extinction toward Pegasus}
It would be useful if the observed photometric properties of the C stars
in Pegasus could shed some light on the exact value of the extinction along
its line of sight. We present, in Table 6, a summary of the photometric
properties of C stars in the program galaxies and the calculated C/M ratios.
Two entries for Pegasus corresponding
to the two currently assumed color excesses are presented. 
If the color excess of Pegasus
is E(R--I) = 0.12, corresponding to E(B--V) = 0.15, we have to select its
C and M stars among stars with (R--I) $>$ 1.02. This shift reduces both
the number of C stars and the number of M stars, thus modifying slightly
 the C/M ratio.

According to Gallagher et al. (1998) Pegasus does not only suffer a 
substantial amount of extinction but is also metal poorer that previously
believed. 
The median color of C stars is known to be a function of the [Fe/H] of the
parent galaxy. From the data of Table 6, we can then investigate how 
 the $<(R-I)_\circ>$ for the two alternatives fits other galaxies. The relevant
data, taken from Paper I and this paper, are displayed in the lower panel of
Figure 6. Pegasus is represented by the two open squares. One sees
that both the high color excess with low metallicity and the low color excess
with larger metallicity fit the relation. If the reddening of Pegasus is high,
its number of C stars would be 30, such number would still fit the relation
between numbers and absolute magnitudes displayed in the upper panel of 
Figure 6. 
\placefigure{fig6}
The AGB models of Groenewegen \& de Jong (1993) show that the mean bolometric
magnitudes of C stars in the Galaxy and in the LMC are essentially identical,
thus possibly independent of the abundance. Current data, on hand, suggest
that the median absolute I magnitude of C stars is fairly 
constant from galaxy to galaxy and is $M_I \approx -4.6$. Our data from
Table 6 along with those assembled in Table 1 lead, for Pegasus, to median
magnitudes of --4.4, for the low reddening, and --4.2 for the high reddening.
The high reddening yields marginally too faint C stars, if the universality
of $<M_I>$ is indeed valid. 

Clues, as to the precise extinction toward Pegasus, must then come from
independent sources and not from C stars.  
Krienke \& Hodge (1998, 2000) 
have investigated colors of background galaxies seen in the field of
Pegasus.  They found that background galaxies suffer 
little reddening when their colors are compared to mean colors of galaxies
as compiled by Buta et al. (1994, 1995). This finding suggests that the color
excess of Pegasus is normal for its galactic latitude. It is indeed low and
Pegasus does have some 40 C stars.
\placefigure{fig7}
The color-magnitude diagrams of Pegasus and the DDO 210 fields are 
displayed in Figure 7. Stars
with $\sigma_{R-I} < 0.3$ are plotted. We see that we have acquired the
first two magnitudes of the giant branch. The C stars 
nicely fit on the extended giant branch. One notices that DDO 210 is at a lower
galactic latitude than Pegasus because of the obvious ridge seen at R--I = 0.4.

\subsection{\it DDO 210}
C02, one of the three C stars of DDO 210, is located 3.2 arcmin from
the center of the galaxy.  At the distance of DDO 210, this corresponds to
nearly 
1 kpc from center of DDO 210. It is however along its major axis and well within
the region where the old population of DDO 210 is found (Lee et al. 1999).
The three stars of DDO 210 are fairly bright when compared to the ones of
Pegasus. 
Could these three stars actually represent the brightest
members of an unobserved large population?  At least two facts suggest 
otherwise. First, if we reasonably relax our acceptance criterion, 
such as allowing
stars with larger color errors to be accepted as C stars, we obviously
can increase the number of carbon star candidates in DDO 210. The added 
candidates however are distributed all over the CCD fields and are much fainter
than the three listed in Table 5. 
They do not represent a slightly fainter population in DDO 210. Second,
the number of C stars in dwarf galaxies
is roughly a function of their mass or absolute magnitude, as we have shown in
Fig. 6. DDO 210 is
among the faintest  dwarf galaxies and it is therefore expected to have few
C stars, possibly 4 C stars according to Fig. 6. 
Thus the C stars in DDO 210 may be unusually bright. Furthermore, DDO 210 
must have very few M giants. Our foreground estimate of M stars is
identical with the total number of M stars seen in the field. According to 
the Groenewegen \& de Jong (1993) synthetic AGB models, the bulk of M stars
is over one magnitude fainter than the C stars, thus a few undetected fainter
ones might be present.
\section{CONCLUSION}
The identification of C stars in a dwarf galaxy points to the presence of
an intermediate age population. We have here presented observations of three
low mass galaxies. Their numbers of C stars are rather  low and 
preclude further detailed analysis at the present time. Data for 
Pegasus and DDO 210 indicate that C stars are certainly not restricted to
the core of these galaxies but extend into their halo. C stars in IC 1613
extend further out than its observed HI (Paper I). The HI maps
of DDO 210 and Pegasus produced  by Lo et al. (1993) show also that the
neutral hydrogen is more concentrated than the C stars. C stars could then
be more useful than HI to investigate the kinematics of 
the periphery  of disk galaxies such as
M31, NGC 55 or WLM. The universality of the median I magnitude of C stars 
appears, following the DDO 210 observations, to be far from certain. One
would definitively need to confirm this with a galaxy with several dozen
C stars. Several luminous Local Group dwarf galaxies need to be
surveyed for C stars.

\acknowledgments
This project is supported financially, in part, by the Natural Sciences and
Engineering Research Council of Canada (S. D.).

\clearpage



\figcaption[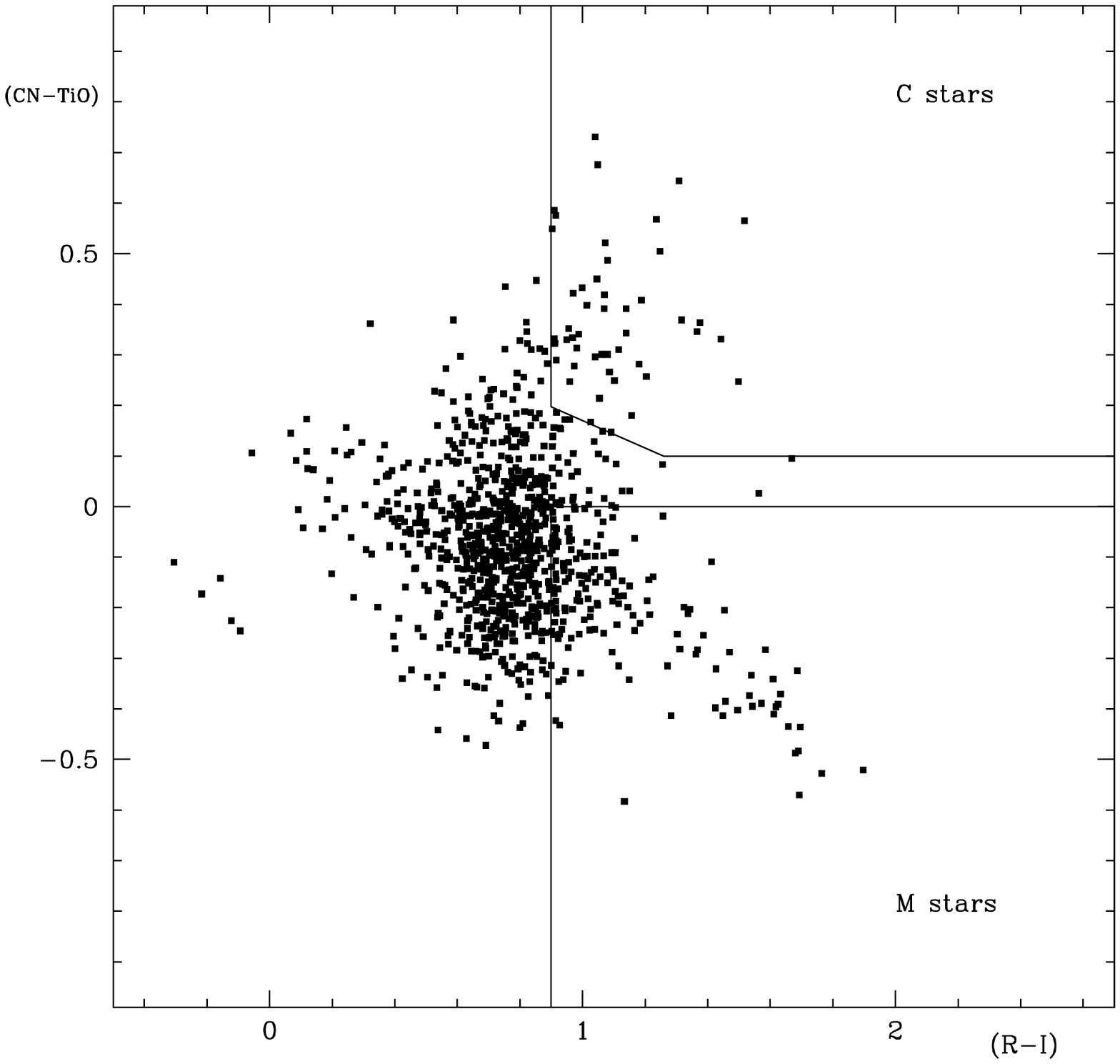]{The color-color diagram of the stars in the Pegasus
field.
The blue limit is drawn at (R--I) = 0.90 but should be adjusted to
take into account the reddening toward Pegasus.\label{fig1}}

\figcaption[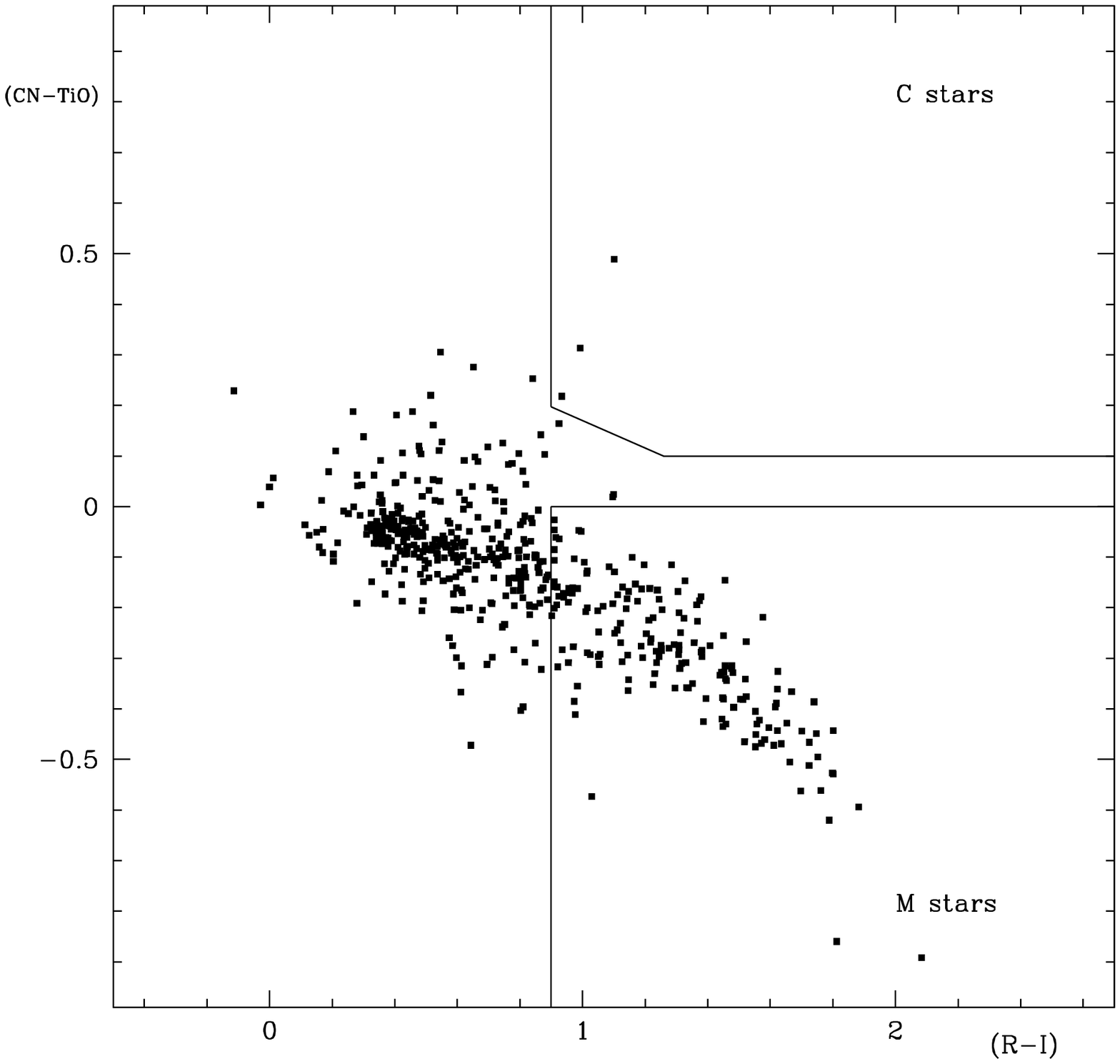]{The color-color diagram of the DDO 210 field.
\label {fig2}}

\figcaption[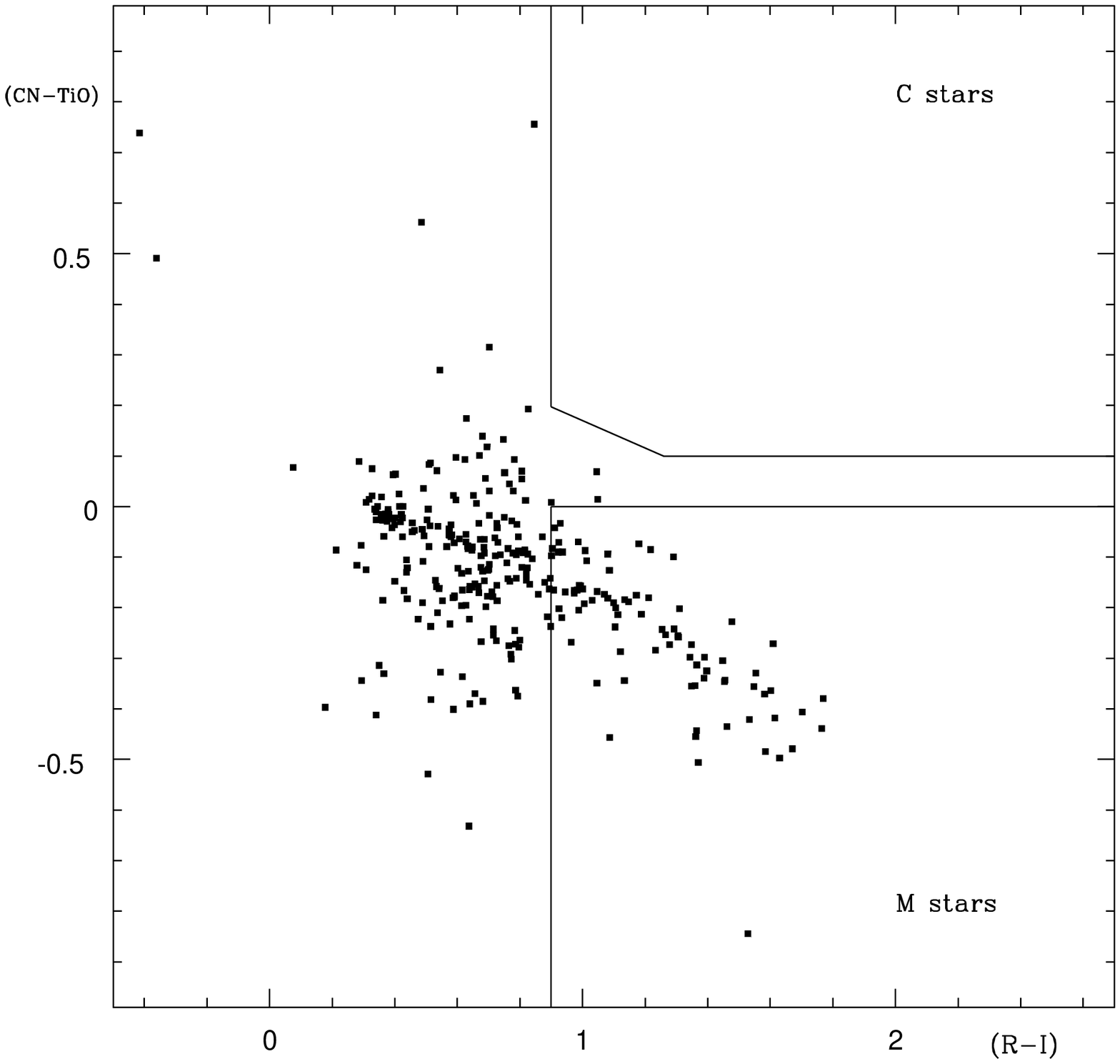]{The color-color diagram of the Tucana field. 
\label{fig3}}

\figcaption[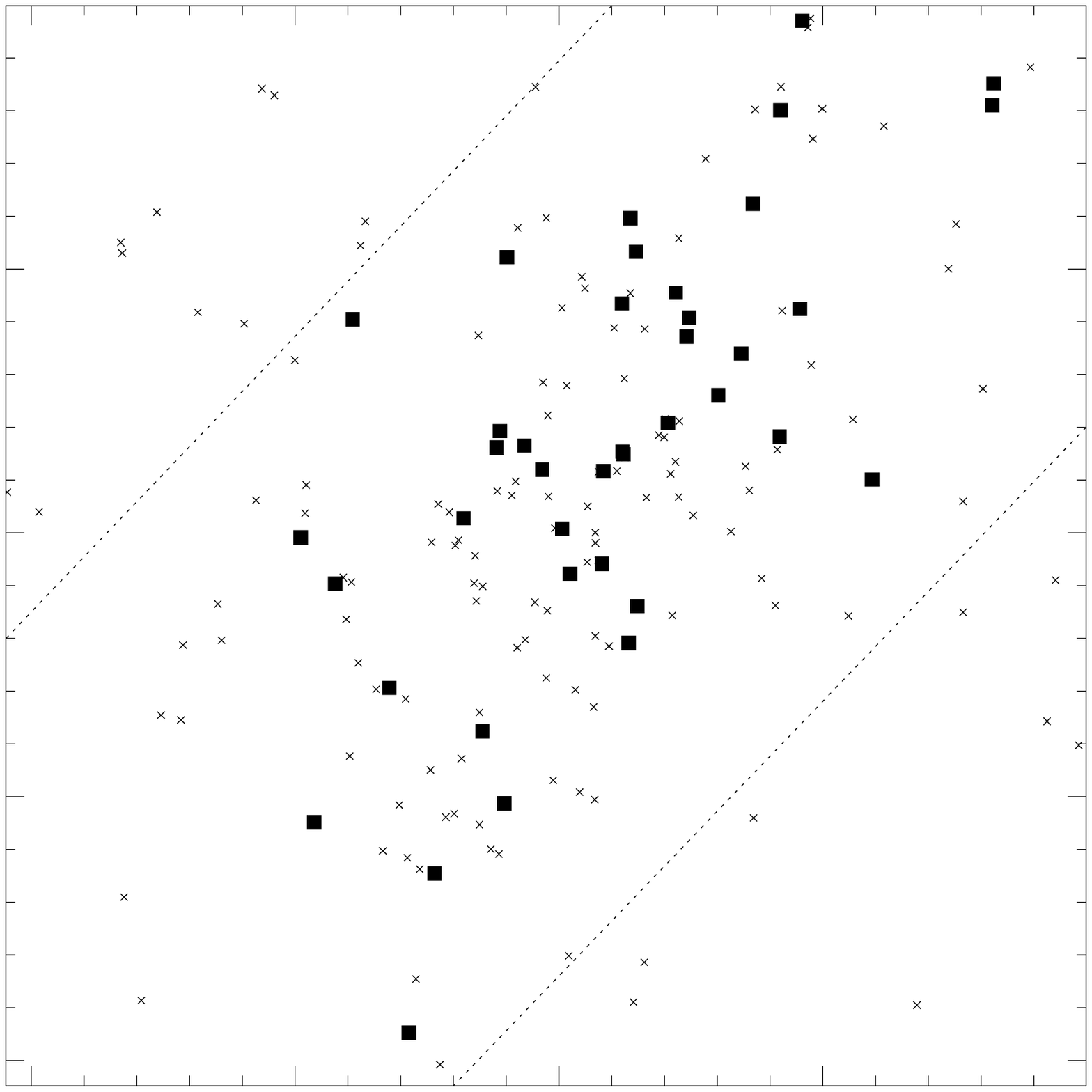]{The spatial distribution of C stars 
(big dots) and M stars (crosses) in the field of Pegasus. The two
dotted lines delineate the ``foreground'' area. North on top, East to the 
left.\label{fig4}}

\figcaption[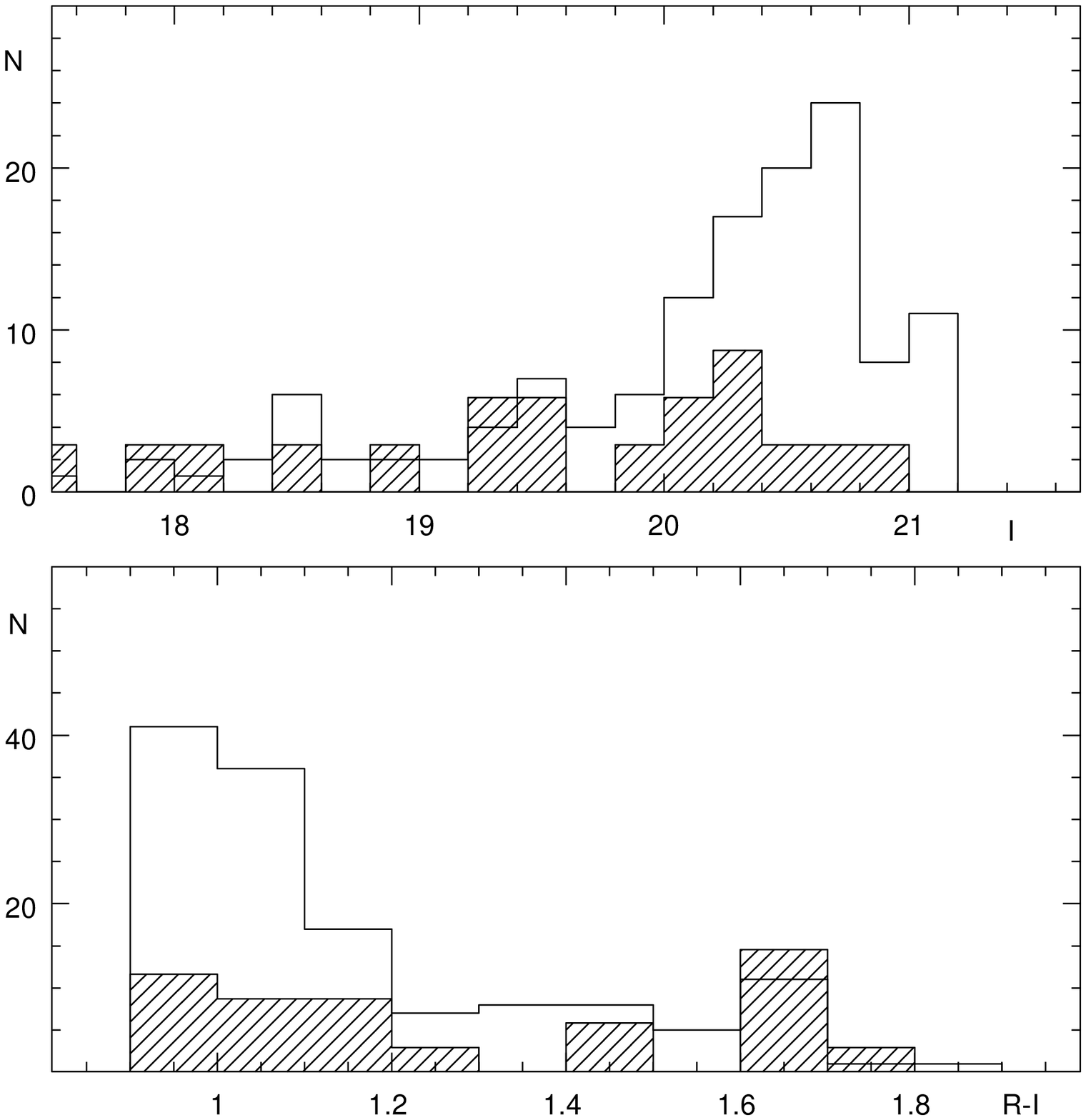]{Comparison between the properties of
M stars in the field of 
Pegasus and the properties of M stars in the foreground. The latter are
represented 
by the shaded histograms. The upper panel shows the magnitude distributions
and the lower panel presents the color distributions.
\label{fig5}}

\figcaption[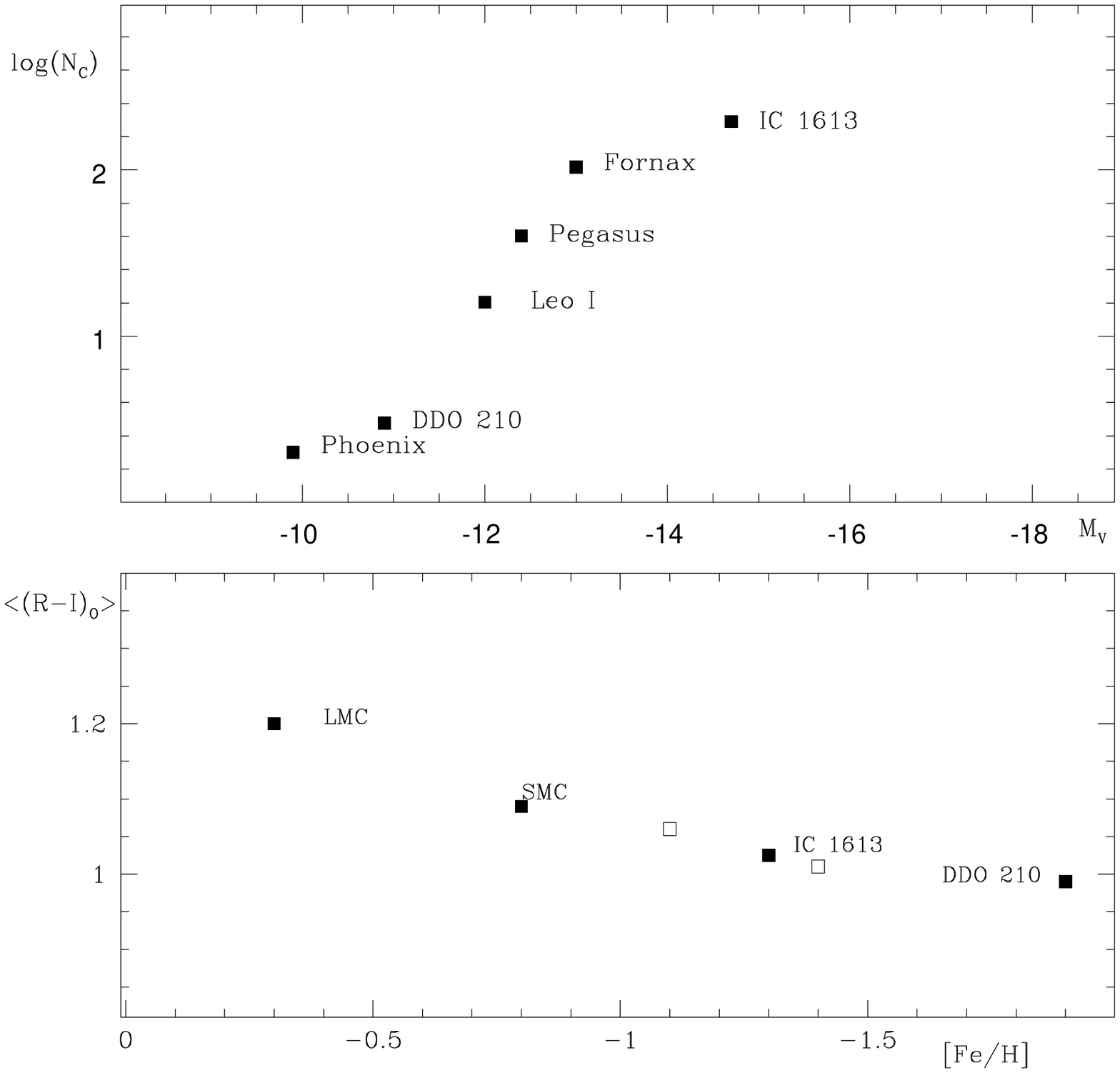] {In the top panel, a plot of the number of C stars, 
in fully surveyed dwarfs, versus the M$_V$ of the parent galaxy, 
shows an obvious trend. In the
lower panel, we display the relationship between the intrinsic median
color of C stars and the [Fe/H] of the parent galaxy. The two open squares
represent Pegasus for E(R--I) = 0.02 (left)  and 0.12 (right) respectively. 
Both points fit the trend seen among galaxies. \label{fig6}}

\figcaption[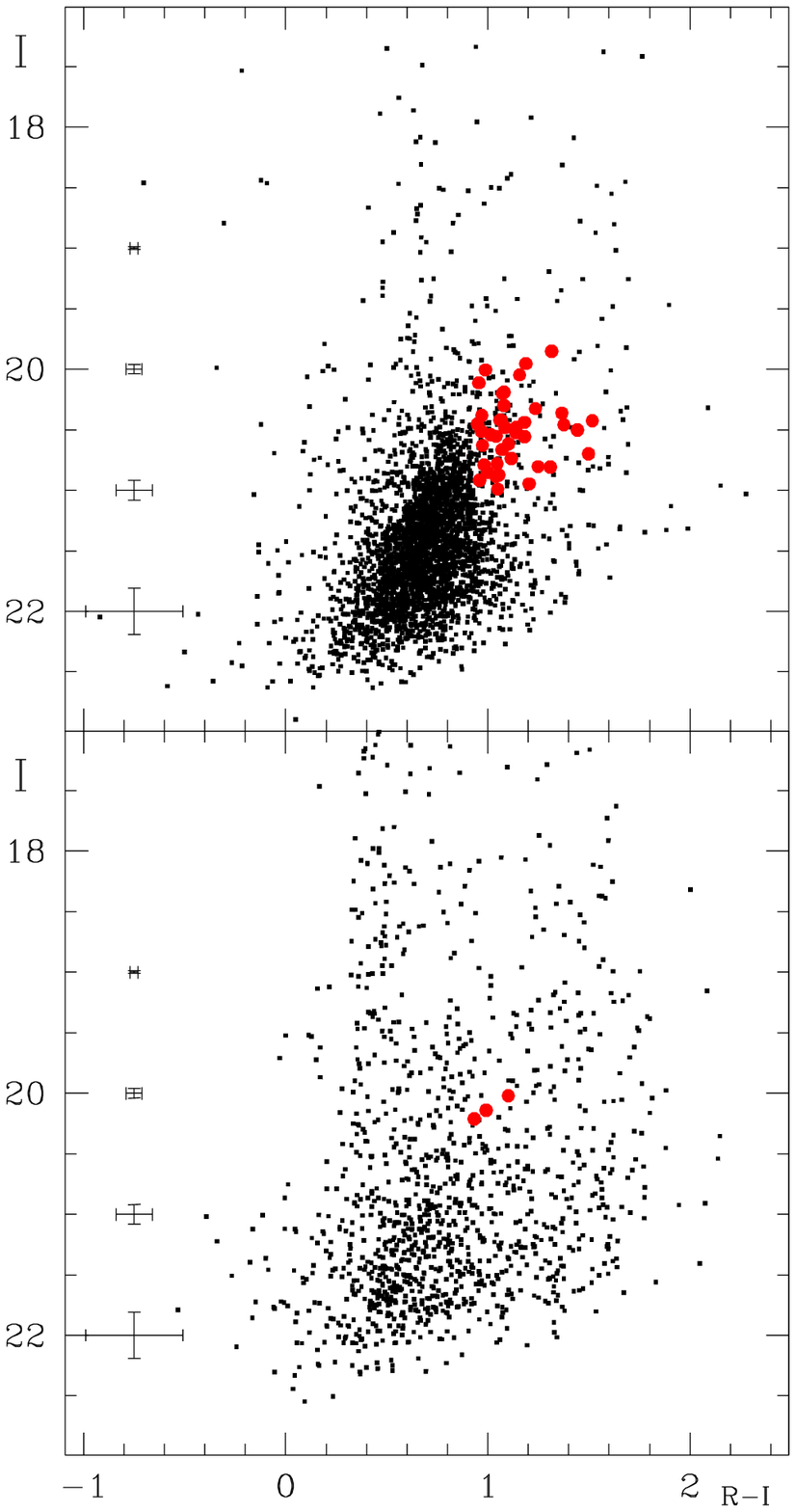]{Color-magnitude diagram of Pegasus (upper panel) and
DDO 210 (lower panel). C stars are represented by big dots. Stars with color
uncertainties $> 0.3$ were excluded, this explains why the limit is $I \approx
22$.  \label{fig7}}
\clearpage





\clearpage





\end{document}